# Van Hove Singularities and Excitonic Effects in the Optical Conductivity of Twisted Bilayer Graphene

Robin W. Havener[†], Yufeng Liang[‡], Lola Brown[§], Li Yang[‡], Jiwoong Park[*,§,∥]

[†]*School of Applied and Engineering Physics, Cornell University, Ithaca, NY 14853, United States*,
[‡]*Department of Physics, Washington University in St. Louis, St. Louis, MO 63130, United States,*
[§]*Department of Chemistry and Chemical Biology, and* [∥]*Kavli Institute at Cornell for Nanoscale Science, Cornell University, Ithaca, NY 14853, United States*

*Correspondence should be addressed to jpark@cornell.edu

**Abstract:** We report a systematic study of the optical conductivity of twisted bilayer graphene (tBLG) across a large energy range (1.2 eV to 5.6 eV) for various twist angles, combined with first-principles calculations. At previously unexplored high energies, our data show signatures of multiple van Hove singularities (vHSs) in the tBLG bands, as well as the nonlinearity of the single layer graphene bands and their electron-hole asymmetry. Our data also suggest that excitonic effects play a vital role in the optical spectra of tBLG. Including electron-hole interactions in first-principles calculations is essential to reproduce the shape of the conductivity spectra, and we find evidence of coherent interactions between the states associated with the multiple vHSs in tBLG.

In two-dimensional materials with van der Waals interlayer coupling, the rotation angle (θ) between layers has emerged as an important degree of freedom with significant effects on these materials' electronic and optical properties. Twisted bilayer graphene (tBLG), a prototypical bilayer system, has been the subject of many recent theoretical and experimental studies[1-6]. In tBLG, the interlayer interactions perturb the band structure of each graphene layer to create new, θ-dependent van Hove singularities (vHSs), which have been observed by scanning tunneling spectroscopy[5,7,8] and optical spectroscopy[9-14]. However, these previous studies focused on relatively low energies where single layer graphene (SLG) has a unique linear band structure. On the contrary, the band structure of SLG becomes more complex at higher energies: the bands lose their linearity, electrons and holes are no longer symmetric, and a saddle point vHS occurs at the M point in the SLG Brillouin zone[15,16]. To date, little is known about how this asymmetry and nonlinearity affects the θ-dependent vHSs and associated optical properties in tBLG.

In addition, while it is known that there are resonant excitons associated with the M point vHS in SLG[17-19], the excitonic effects associated with the interlayer vHSs in tBLG are poorly understood. Furthermore, new excitonic states could form as coherent combinations of the multiple intralayer and interlayer vHSs in tBLG, particularly those closest in energy. Understanding these effects in tBLG would also aid our understanding of other stacked and twisted two-dimensional materials, such as hexagonal boron nitride and transition metal dichalcogenides, whose single layers have intrinsic band gaps and other higher energy vHSs.

In this work, we perform optical absorption spectroscopy of tBLG with known θ to explore the tBLG band structure and many-body states. For the first time, we present the full optical absorption spectra of tBLG with various θ over a large energy range (1.2 – 5.6 eV), which encompasses multiple θ-dependent vHSs in tBLG as well as the absorption peak associated with the M point vHS in SLG. We find that the energies of the features we observe correspond to those of the θ-dependent vHSs. However, the detailed absorption spectra we measure differ significantly, particularly at large θ, from those calculated using a tight binding model[20]. Instead, we find that new first-principles calculations which account for electron-hole interactions are needed to describe our data. Moreover, we observe signatures of coherence between the multiple vHSs in tBLG, which have similar energies at large θ, in the form of the enhancement of the lower energy optical response and suppression of the higher energy features.

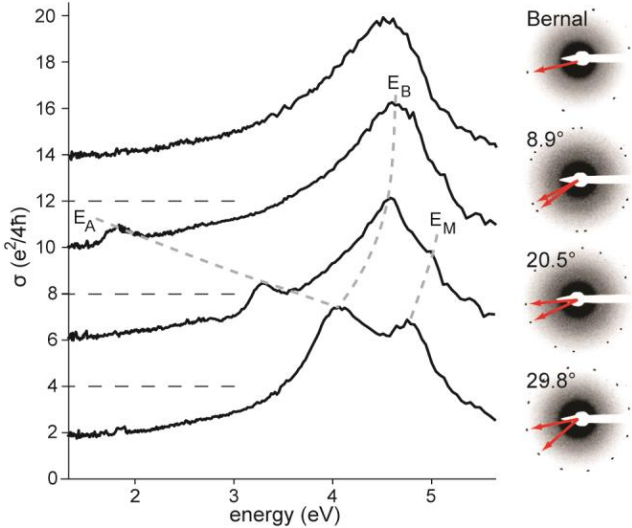

**Figure 1:** (left) Measured σ spectra of BLG (top), and tBLG with increasing θ. Dashed lines are guides to the eye indicating three features in the σ spectra of tBLG which are not found in BLG. Spectra are offset in σ by $4e^2/4\hbar$ apiece for clarity. (right) Electron diffraction patterns (inverted and contrast enhanced) from the same bilayer regions, indicating θ for each.

Figure 1 summarizes our main experimental results. Using a specialized DUV-Vis-NIR hyperspectral microscope with sub-micron spatial resolution[14], we have extracted the optical conductivity spectra for 34 chemical vapor deposition grown tBLG domains (typically a few microns or larger in size) with 8° < θ < 30°. The twist angle of these domains is known to within a fraction of a degree after electron diffraction measurements of the same samples, which sit on electron-transparent, 10 nm thick silicon nitride windows[14]. While we obtain the full complex optical conductivity (see Supporting Information) from a combination of reflection and transmission spectroscopy, we focus here on its real part, referred to in this manuscript as σ. The σ spectrum of tBLG shown in Figure 1 exhibits a number of θ-dependent features. Compared with Bernal stacked bilayer graphene (BLG) (top curve), which has a constant $\sigma = 2e^2/4\hbar$ at infrared and visible energies[21] and a peak due to its saddle point exciton near 4.6 eV[17,18], an extra σ peak ($E_A$) appears in tBLG whose energy increases from ~2 to 4 eV with increasing θ. While the relationship between θ and the energy of this peak has been the subject of several previous studies[10,11,13,14], the full σ spectrum of tBLG has not been presented for energies greater than 3 eV. Additionally, at higher energies (~4.5-5 eV), we find that the σ spectrum is also strongly modified compared to that of BLG. Most noticeable is the additional dip and peak which grows in magnitude as θ increases, significantly modifying the σ spectrum of tBLG near 30º. This is due to the presence of two new peaks, $E_B$ and $E_M$, which we identify based on our analysis in the following Figures 2 and 4, respectively.

To more clearly distinguish small changes in σ between tBLG and BLG, particularly the high energy features that are difficult to resolve over the large absorption peak at 4.6 eV, we subtract the σ of BLG ($\sigma_B$) from that of tBLG ($\sigma_T$). We first examine the general trends in all of our collected $\sigma_T - \sigma_B$ spectra by combining them into a two-dimensional plot, as shown in Figure 2a. Both the $E_A$ and $E_B$ features indicated in Figure 1 are clearly visible at all angles, and the two peaks each lie on single, monotonic curves (solid lines denoted as $E_A$ and $E_B$, Figure 2a) in energy vs. θ.

These two peaks correspond to the interlayer vHSs in tBLG bands[1,20]. Briefly, the band structure of tBLG can be viewed as two sets of SLG bands with a relative rotation of θ around the Γ point (Figure 2a, inset). Interlayer interactions perturb the bands from each layer where they intersect; this occurs along two distinct lines, whose projections onto the momentum plane are labeled $I_A$ and $I_B$ (Figure 2b). The eigenstates near these intersections are then hybridized between both layers, producing a minigap (Δ) and new vHSs. The allowed optical transitions near $I_A$ are those shown in Figure 2c[11,20,22]. Importantly, for a Dirac band structure, both of the allowed transitions have the same energy, $E_A$. In the simplified picture in Figure 2c, $E_A$ is independent of Δ at the band intersection and can be calculated based on the band energies of SLG for a given θ.

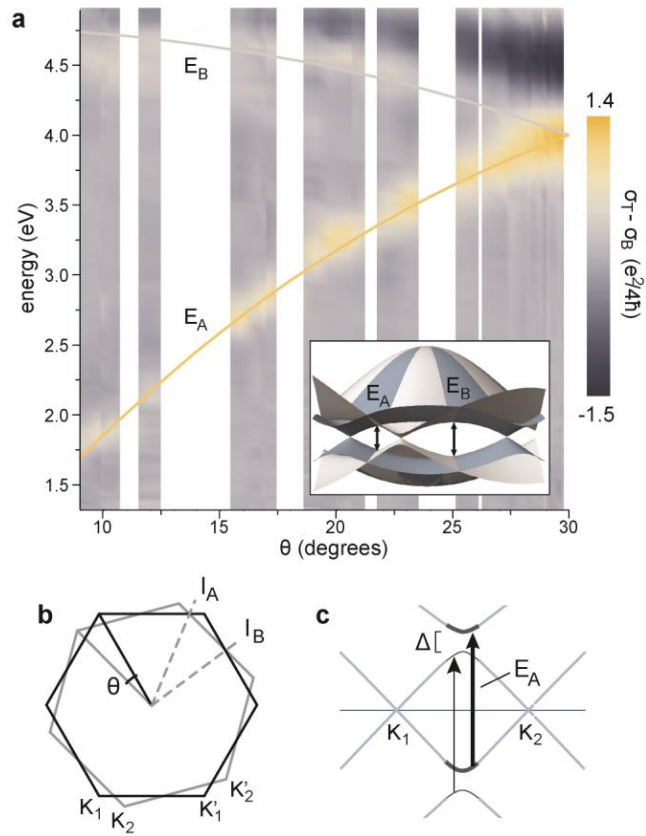

**Figure 2:** (a) A 2D plot of $\sigma_T - \sigma_B$ combining all spectra. A linear background was subtracted from each spectrum after averaging every 10 nm. The two main features, $E_A$ and $E_B$, are fit to the model described in the text (inset). (b) Brillouin zones of each tBLG layer, rotated by θ. The lines $I_A$ and $I_B$ indicate where the bands from each layer intersect. (c) tBLG bands along the direction crossing $K_1$ and $K_2$ and perpendicular to $I_A$. At $I_A$, a minigap opens with energy Δ. The allowed optical transitions near $I_A$ are indicated by thick and thin arrows, which have a constant magnitude $E_A$ between states from parallel bands.

The applicability of this simple picture, similar to the established continuum model of tBLG[1], has not been experimentally confirmed to date for energies > 4 eV. Our data show that both $E_A$ and $E_B$, associated with the interlayer vHSs located along $I_A$ and $I_B$, respectively, can be predicted using this model for θ up to 30° assuming the selection rules shown in Figure 2c and the known SLG band structure[16]. We obtain a remarkably close fit (solid lines, Fig. 2b) to both peaks after a 4% increase in energy (see Supporting Information; similar analysis was performed in Ref. 14 for $E_A$ only). Furthermore, our data provide a facile, all-optical identification method for θ in future samples. For convenience, we fit $E_A$ to an empirical function $\theta = A - (B - CE_A)^{1/2}$, with θ in degrees and $E_A$ in eV. We find A = 41.4, B = 1.74×10³, and C = 3.99×10² eV⁻¹ fit our data for 8° < θ < 30° with a slightly smaller total mean squared error than the fit in Figure 2a.

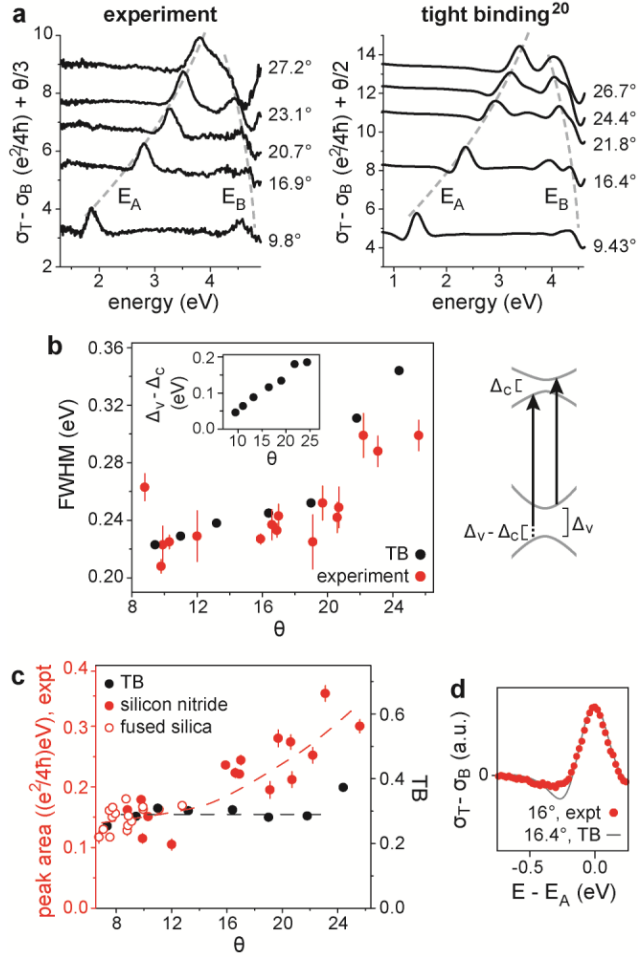

**Figure 3:** (a) Experimental and TB (adapted from Ref. 20; see main text) $\sigma_T - \sigma_B$ spectra at similar angles, with offsets proportional to θ. Note different scales in each plot. (b, left) Peak FWHM, which increases with increasing θ. (inset) TB calculated $E_A$ peak splitting as a function of θ. (b, right) Schematic showing *e-h* asymmetry, which leads to $E_A$ peak broadening by $\Delta_v - \Delta_c$. (c) $E_A$ peak area, determined by a Gaussian fit, as a function of θ. Filled red circles are from spectra shown in Figure 2b, on silicon nitride substrates, and empty circles are from additional samples on fused silica. Filled black circles are areas calculated from TB spectra (note different scale). Dashed lines are guides to the eye. (d) Calculated $\sigma_T - \sigma_B$ at 16.4° (line), alongside averaged experimental data (circles). Four spectra within 2° of 16° were averaged as a function of $E - E_A$, and the calculations and experimental data were scaled to match in height.

Next, we focus on the magnitude and lineshapes of our σ data, plotting several individual $\sigma_T - \sigma_B$ spectra in Figure 3a. We compare these results to recent tight binding (TB) calculations of $\sigma_T$ at similar commensurate angles[20], after subtracting the calculated $\sigma_B$ from Ref. 20 and broadening the result by Gaussian convolution ($\sigma_{Gauss}$ = 0.11 eV). The empirical broadening needed to match our data is much larger than our experimental resolution (2 nm, or ~0.015 eV at 3 eV), indicating additional thermal and/or inhomogeneous broadening of the optical response in our samples. The TB calculations capture several features we observe. First, they clearly show the presence of the $E_A$ and $E_B$ peaks whose positions and shapes evolve with θ. Second, we observe that the $E_A$ peak broadens as θ increases. Figure 3b plots the FWHM of a Gaussian fit vs. θ for all θ < 27° (red dots). The TB widths are numerically consistent with our data (Fig. 3b, solid dots) and suggest that the broadening is caused by the band asymmetry between electrons and holes at larger energies, discussed in Ref. 20 (Figure 3b, right), which originates from a non-zero overlap integral in the tight binding picture of the graphene band structure[23]. Here, electron-hole asymmetry alters the magnitudes of the minigaps near $I_A$ in the valence ($\Delta_v$) and conduction ($\Delta_c$) bands, splitting the tBLG absorption feature into two peaks with separation $\Delta_v - \Delta_c$. This splitting, which increases with θ up to ~0.2 eV near 25° (inset, Figure 3b), is too small to be seen directly in our data; instead, it contributes to peak broadening.

However, the TB results deviate from experiment in several ways. First, there are significant differences between the calculated and observed peak shapes, especially for θ > 20°, as can be seen in Figure 3a. In particular, the relative weights of the $E_A$ and $E_B$ features do not match our results. Second, the relative areas of the $E_A$ peaks as a function of θ are also not well described by the TB calculations. Figure 3c shows the integrated area of the $E_A$ peak, determined by a Gaussian fit. For this plot, we included the σ spectra of 12 additional tBLG domains on fused silica substrates using reflection spectroscopy[17], converting $E_A$ to θ with our empirical formula. Our experimental results show an approximately constant peak area at low angles, followed by an increase at higher angles. On the contrary, the TB calculated area is roughly constant for 10° < θ < 20°. Finally, we observe a subtle, but significant, difference in the $E_A$ peak shape between theory and experiment. Figure 3d compares the lineshapes of the $E_A$ peak calculated for 16.4° with the experimental data averaged for samples near θ ~ 16°. The dip before the $E_A$ peak is more pronounced in the TB calculation than it is in our data.

These differences may be caused by excitonic interactions in tBLG, which were not included in the single particle TB calculations. In fact, deviations from the single-particle prediction of the optical absorption lineshape in other low-dimensional carbon materials, such as SLG[17-19] and metallic SWNTs[24], have been attributed to excitonic effects. Little is currently known about the effects of electron-hole (*e-h*) interactions on the optical properties of tBLG. To study this further, we compute the optical absorption spectrum of tBLG using first-principles GW-Bethe-Salpeter-Equation (BSE) simulations[25-28]. We consider the two angles which possess the smallest two primitive cells: 21.8° and 27.8°, with 28 and 52 carbon atoms, respectively. To obtain the absorption features over a large spectral range and to guarantee a converged optical spectrum with excitonic effects included, we use a denser k-point-sampling over the first Brillouin zone than previous work[29] (60×60×1 for 21.8° and 36×36×1 for 27.8°) and include several conduction and valence bands in the calculation (7 bands each for 21.8° and 12 each for 27.8°; see Supporting Information for more details).

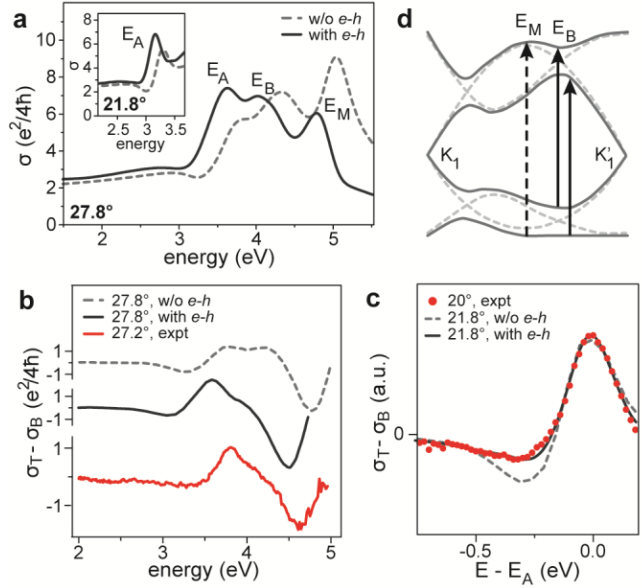

**Figure 4:** (a) Calculated σ spectra of tBLG with and without *e-h* interactions included. The *e-h* interactions act to redshift the absorption features, and increase the spectral weight of the lower energy peaks while decreasing that of the higher energy peaks. (b) Calculated $\sigma_T - \sigma_B$ at 27.8° alongside data at 27.2°. The *e-h* calculation captures the relative heights of the peaks seen in experiment (note different *y* scales). (c) Calculated $\sigma_T - \sigma_B$ at 21.8° (lines), alongside averaged experimental data (circles). (d) $E_B$ and $E_M$ optical transitions are overlaid on tBLG band structure without (dashed lines) and with (solid lines) interlayer interactions included. The $E_M$ transition is weakened because the selection rules for the nearby $E_B$ transition are different.

Our calculations indicate that excitonic effects play an important role in the optical response of tBLG. Figure 4a contrasts the calculated tBLG absorption spectra with and without *e-h* interactions. The inclusion of *e-h* interactions lowers $E_A$ by ~200 meV for both angles because of the reduced screening between quasiparticles. While these shifts are smaller than the exciton binding energy of 600 meV estimated for the saddle-point exciton in SLG[19], they are an order of magnitude larger than those predicted in metallic SWNTs[30,31].

In addition, our GW-BSE calculations reproduce several key experimental features that the TB calculations fail to explain, and suggest that the excitonic effects on the optical absorption spectra of tBLG are qualitatively different from those found in SLG and BLG. First, the spectral weight of the absorption curve redshifts after *e-h* interactions are included. The lowest energy $E_A$ peak experiences the largest enhancement, the $E_B$ peak is roughly unchanged, and the highest energy peak associated with the M point vHS of SLG ($E_M$) is substantially reduced. This trend differs significantly from what is observed in SLG or BLG, where excitonic effects are less pronounced at lower energies. The difference in tBLG is that the energies of the interlayer vHSs are close to each other for large θ. Because of this, their excitonic interaction energy regimes overlap and make it possible for distant interband transitions to contribute optical oscillator strength to lower-energy excitons, such as $E_A$. Therefore, the optical absorbance of the lower-energy excitons is enhanced in tBLG. This behavior is consistent with our data in Figure 3c: at higher angles, where the vHSs $E_A$ and $E_B$ are the closest in energy, the $E_A$ peak area is enhanced the most strongly with respect to that predicted by TB calculations.

Moreover, these shifts in spectral weight provide a close match to the experimental lineshape of σ in tBLG. In Figure 4b, we plot the calculated $\sigma_T - \sigma_B$ at 27.8°, with and without *e-h* interactions, alongside our data at a similar angle (27.2°). The overall shape of the calculated spectrum with *e-h* interactions is a much better fit, especially the relative weights of the $E_A$ and $E_B$ features. Finally, the *e-h* calculation provides a closer match to the $E_A$ peak shape we observe. Figure 4c, comparing the experimental $E_A$ peak near θ ~ 20° to the calculations at 21.8° (similar to Fig 3d), shows that the dip below the absorption peak becomes less pronounced after *e-h* interactions are included.

So far, we have focused on the two lower energy peaks in the tBLG spectrum, $E_A$ and $E_B$. However, the interlayer vHSs also strongly perturb the peak associated with graphene's intrinsic saddle point vHs, $E_M$. In SLG and BLG, the $E_M$ peak dominates the optical absorption spectrum (see Figure 1), and its exciton-induced red shift is calculated to be ~600 meV[19]. However, in tBLG, we find that the excitonic red shift of the $E_M$ peak is only 210 meV for 27.8° (Figure 4a). In addition, the $E_M$ peak height is calculated to be ~$5e^2/4\hbar$, about half that found in BLG. Our experimental results in Figure 1 qualitatively reproduce the reduced height and blueshift of the $E_M$ peak at high angles. Further analysis of our calculations shows that the screened Coulomb interactions in BLG and tBLG are not very different. However, we find that the M point optical transition strength is reduced by the $E_B$ transition. Since the intralayer optical transition at the M point occurs between a different pair of bands than either of the $E_B$ transitions (Figure 4d), these transitions compete when they are close to each other in *k*-space, weakening the M point transition compared to that in BLG. Both the spectral weight redshift discussed above and the weakening of the intralayer transitions at the M point likely cause the reduction of the excitonic effects near the $E_M$ peak in tBLG.

A few discrepancies remain between our experimental data and both TB and first principles calculations. First, all of the calculated values of the absolute magnitudes of the features in $\sigma_T - \sigma_B$ are significantly larger than experiment, which can be seen from the different y-axis scales used for the theory results in Figures 3 and 4. Second, the peak positions calculated with GW-BSE are significantly lower than the experimental results. Both of these discrepancies may reflect overestimated interlayer coupling. The local density approximation (LDA) is the starting point for all of the DFT calculations in this manuscript (the TB parameters for tBLG are fit to DFT band structure calculations), and this approximation can overestimate the degree of interlayer hopping in bilayer graphene, resulting in vHSs of a larger magnitude and stronger excitonic effects after the GW-BSE framework is included. While DFT has successfully modeled BLG, there is no guarantee that this technique will accurately capture the weaker interlayer coupling in tBLG. Several additional factors may influence the reduced experimental peak intensity in comparison to the theoretical predictions. Charge inhomogeneities, which are common in 2D materials[32], could reduce the strength of the σ features we measure. Specifically, small potential differences across the layers would shift their valence and conduction band intersections to different locations in *k*-space[1], which would reduce the allowed optical transitions between them. Local doping could also modify the many-body contributions to σ by modifying screening. Finally, thermal fluctuations may also affect the tBLG coupling strength[11]. Further studies will be needed to examine each of these effects in more detail.

In conclusion, we have characterized the optical conductivity of tBLG over a large energy and angle range. The energies of the features associated with the interlayer vHSs can be fit to a simple model, but the shapes of our absorption peaks are matched best by calculations which include excitonic effects. Our results provide a framework for modeling the effects of interlayer interactions on the properties of not only tBLG, but also vertical heterostructures of a variety of other materials, such as hexagonal boron nitride and transition metal dichalcogenides. The interactions between twisted layers provide a further degree freedom which, if controlled, could potentially be exploited to tune the properties of all stacked two dimensional materials.

**References**
1. dos Santos, J.; Peres, N.; Castro, A. *Phys. Rev. Lett.* **2007**, *99*, 256802.
2. Mele, E. *Phys. Rev. B* **2010**, *81*, 161405.
3. Shallcross, S.; Sharma, S.; Kandelaki, E.; Pankratov, O. *Phys. Rev. B* **2010**, *81*, 165105.
4. Bistritzer, R.; MacDonald, A. *PNAS* **2011**, *108*, 12233-12237.
5. Li, G.; Luican, A.; dos Santos, J.; Neto, A.; Reina, A.; Kong, J.; Andrei, E. *Nat. Phys.* **2010**, *6*, 109-113.
6. Ohta, T.; Robinson, J. T.; Feibelman, P. J.; Bostwick, A.; Rotenberg, E.; Beechem, T. E. *Phys. Rev. Lett.* **2012**, *109*, 186807.
7. Luican, A.; Li, G.; Reina, A.; Kong, J.; Nair, R.; Novoselov, K.; Geim, A.; Andrei, E. *Phys. Rev. Lett.* **2011**, *106*, 126802.
8. Brihuega, I.; Mallet, P.; Gonzalez-Herrero, H.; de Laissardiere, G.; Ugeda, M.; Magaud, L.; Gomez-Rodriguez, J.; Yndurain, F.; Veuillen, J. *Phys. Rev. Lett.* **2012**, *109*, 196802.
9. Ni, Z.; Liu, L.; Wang, Y.; Zheng, Z.; Li, L.; Yu, T.; Shen, Z. *Phys. Rev. B* **2009**, *80*, 125404.
10. Wang, Y.; Ni, Z.; Liu, L.; Liu, Y.; Cong, C.; Yu, T.; Wang, X.; Shen, D.; Shen, Z. *ACS Nano* **2010**, *4*, 4074-4080.
11. Havener, R. W.; Zhuang, H.; Brown, L.; Hennig, R. G.; Park, J. *Nano Lett.* **2012**, *12*, 3162-3167.
12. Kim, K.; Coh, S.; Tan, L.; Regan, W.; Yuk, J.; Chatterjee, E.; Crommie, M.; Cohen, M.; Louie, S.; Zettl, A. *Phys. Rev. Lett.* **2012**, *108*, 246103.
13. Robinson, J. T.; Schmucker, S. W.; Diaconescu, C. B.; Long, J. P.; Culbertson, J. C.; Ohta, T.; Friedman, A. L.; Beechem, T. E. *ACS Nano* **2012**, *7*, 637-644.
14. Havener, R. W.; Kim, C. J.; Brown, L.; Kevek, J. W.; Sleppy, J. D.; McEuen, P. L.; Park, J. *Nano Lett.* **2013**, *13*, 3942.
15. Reich, S.; Maultzsch, J.; Thomsen, C.; Ordejon, P. *Phys. Rev. B* **2002**, *66*, 035412.
16. Gruneis, A.; Attaccalite, C.; Wirtz, L.; Shiozawa, H.; Saito, R.; Pichler, T.; Rubio, A. *Phys. Rev. B* **2008**, *78*, 205425.
17. Mak, K.; Shan, J.; Heinz, T. *Phys. Rev. Lett.* **2011**, *106*, 046401.
18. Chae, D.; Utikal, T.; Weisenburger, S.; Giessen, H.; von Klitzing, K.; Lippitz, M.; Smet, J. *Nano Lett.* **2011**, *11*, 1379-1382.
19. Yang, L.; Deslippe, J.; Park, C.; Cohen, M.; Louie, S. *Phys. Rev. Lett.* **2009**, *103*, 186802.
20. Moon, P.; Koshino, M. *Phys. Rev. B* **2013**, *87*, 205404.
21. Nair, R.; Blake, P.; Grigorenko, A.; Novoselov, K.; Booth, T.; Stauber, T.; Peres, N.; Geim, A. *Science* **2008**, *320*, 1308-1308.
22. Tabert, C.; Nicol, E. *Phys. Rev. B* **2013**, *87*, 121402.
23. Saito, R.; Dresselhaus, G.; Dresselhaus, M. S. *Physical Properties of Carbon Nanotubes*; Imperial College Press: London, **1998**.
24. Wang, F.; Cho, D.; Kessler, B.; Deslippe, J.; Schuck, P.; Louie, S.; Zettl, A.; Heinz, T.; Shen, Y. *Phys. Rev. Lett.* **2007**, *99*, 227401.
25. Kohn, W.; Sham, L. *Phys. Rev.* **1965**, *140*, A1133-A1138.
26. Hybertsen, M.; Louie, S. *Phys. Rev. B* **1986**, *34*, 5390-5413.
27. Rohlfing, M.; Louie, S. *Phys. Rev. B* **2000**, *62*, 4927-4944.
28. Deslippe, J.; Samsonidze, G.; Strubbe, D.; Jain, M.; Cohen, M.; Louie, S. *Comput. Phys. Commun.* **2012**, *183*, 1269-1289.
29. Chen, Z.; Wang, X. *Phys. Rev. B* **2011**, *83*, 081405.
30. Spataru, C.; Ismail-Beigi, S.; Benedict, L.; Louie, S. *Phys. Rev. Lett.* **2004**, *92*, 077402.
31. Deslippe, J.; Spataru, C.; Prendergast, D.; Louie, S. *Nano Lett.* **2007**, *7*, 1626-1630.
32. Martin, J.; Akerman, N.; Ulbricht, G.; Lohmann, T.; Smet, J. H.; Von Klitzing, K.; Yacoby, A. *Nat. Phys.* **2008**, *4*, 144-148.
33. Giannozzi, P.; Baroni, S.; Bonini, N.; Calandra, M.; Car, R.; Cavazzoni, C.; Ceresoli, D.; Chiarotti, G.; Cococcioni, M.; Dabo, I.; *et al*. *J. Phys.: Condens. Matter* **2009**, *21*, 395502.

**Acknowledgements**
We thank M. W. Graham, K. F. Mak, and P. L. McEuen for useful discussions. This work was mainly supported by the AFOSR (Grants FA9550-09-1-0691, FA9550-10-1-0410) and the Samsung Advanced Institute for Technology GRO Program, as well as the Kavli Institute at Cornell for Nanoscale Science (KIC). Additional funding was provided by the Nano Material Technology Development Program through the National Research Foundation of Korea (NRF) funded by the Ministry of Science, ICT, and Future Planning (2012M3A7B4049887). This work made use of the Cornell Center for Materials Research Shared Facilities which are supported through the NSF MRSEC program (DMR-1120296). Device fabrication was performed at the Cornell NanoScale Facility, a member of the National Nanotechnology Infrastructure Network, which is supported by the NSF (Grant ECS-0335765). Y. Liang and L. Yang are supported by NSF Grant No. DMR-1207141. The computational resources have been provided by Lonestar of Teragrid at the Texas Advanced Computing Center (TACC). The ground-state calculation was performed by the Quantum Espresso[33], and the GW-BSE calculation was done with the BerkeleyGW package[28].